\documentclass[showpacs,aps,pra,amsmath,amssymb,twocolumn]{revtex4-1}
\usepackage{graphicx,graphics}
\usepackage{dcolumn}
\usepackage{bm,color}
\usepackage{txfonts,hyperref}
\usepackage{amsmath}
\usepackage{epstopdf}
\newcommand{\ket}[1]{\left\vert#1\right\rangle}
\newcommand{\bra}[1]{\left\langle#1\right\vert}

\newcommand{\eq}{Eq.~}
\newcommand{\eqs}{Eqs.~}
\newcommand{\fig}{Fig.~}
\newcommand{\figs}{Figs.~}
\newcommand{\cf} {cf.~}
\newcommand{\ug} {\!=\!}

\newcommand{\piu} {\!+\!}
\newcommand{\meno} {\!-\!}

\newcommand{\eg} {e.g.~}

\newcommand{\rref} {Ref.~}
\newcommand{\rrefs} {Refs.~}
\newcommand{\vs} {vs.~}

\newcommand{\ibid}{{\it ibid.~}}

\providecommand{\ket}[1]{|#1\rangle}
\providecommand{\bra}[1]{\langle#1|}
\providecommand{\kebra}[2]{\ket{#1}\!\bra{#2}}

\begin{document}

\pacs{03.65.Yz, 03.67.--a}

\title{Non-Markovianity of a quantum emitter in front of a mirror}

 \author{Tommaso Tufarelli \mbox{$^{1}$}}
\author{M. S. Kim\mbox{$^{1}$}}
\author{Francesco Ciccarello\mbox{$^{2}$}}

\affiliation{\\ 
\mbox{$^{1}$QOLS, Blackett Laboratory, Imperial College London, SW7 2BW, UK}\\\mbox{$^{2}$NEST, Istituto Nanoscienze-CNR and Dipartimento di Fisica, Universit$\grave{a}$  degli Studi di Palermo, via Archirafi 36, I-90123 Palermo, Italy}}

\begin{abstract}
We consider a quantum emitter (``atom") radiating in a one-dimensional (1D) photonic waveguide in the presence of a single mirror, resulting in a delay differential equation for the atomic amplitude. We carry out a systematic analysis of the non-Markovian (NM) character of the atomic dynamics in terms of refined, recently developed notions of quantum non-Markovianity such as indivisibility and information back-flow. NM effects are quantified as a function of the round-trip time and phase shift associated with the atom-mirror optical path. We find, in particular, that unless an atom-photon bound state is formed a {\it finite} time delay is always required in order for NM effects to be exhibited. This identifies a finite threshold in the parameter space, which separates the Markovian and non-Markovian regimes.
\end{abstract}

\maketitle
\noindent

\section{Introduction}  
The distinction between Markovian and non-Markovian regimes has long been considered a basic one in the study of open system dynamics, i.e., when the system of interest is in contact with an external environment. In qualitative terms, Markovianity is typically associated with the lack of {\it memory effects}, a situation which considerably simplifies the theoretical description and yet occurs with good approximation in a number of cases. Assessing the (non-)Markovianity of an open dynamics is a well-understood problem in classical mechanics. In the quantum realm -- quite differently -- it is not \cite{petruccione,huelga,breuer}. Until recently, Markovianity has been almost ubiquitously identified with regimes where the open quantum dynamics is well-described by a master equation (ME) of the Kossakowski-Lindblad form \cite{petruccione} --- ``Lindbladian" dynamics for brevity---. The latter typically gives rise to purely exponential decays of quantities such as mean energy, po!
 pulations and coherences. A vast and variegated literature, indeed, has used and in some cases still uses the term ``non-Markovian" as a synonym of non-Lindbladian.

Over the last few years, however, a considerable amount of work has been devoted to the refinement of the very notion of non-Markovianity (NM) of a quantum dynamics, with the aim of providing its rigorous identification and quantification \cite{breuer}. Several new definitions of NM have been proposed, each associated with a specific quantitative measure \cite{measures,rivas,BLP,tore}. A particularly intuitive one is the so called ``BLP" measure \cite{BLP}, which identifies NM with the occurrence of {\it quantum} information (QI) back-flow between the system and environment (i.e., there exist times at which the latter is able to return QI to the former). To appreciate how these recent studies are affecting the pre-existing paradigm of NM, suffice it to say that certain well-known integro-differential MEs were shown to have zero BLP measure \cite{laura-breuer}, despite for a long time a ME of this sort had been regarded as a typical NM process. The analysis of NM from this re!
 newed perspective has been recently applied to a number of systems such as atoms in lossy cavities \cite{lossy1,lossy2}, spin-boson models \cite{clos}, spin chains \cite{tony} and ultracold atoms \cite{pinja}. A major motivation to explore different physical scenarios is that studying the emergence of NM in different environmental models helps our understanding of NM itself, a concept whose physical meaning is currently under debate \cite{debate}. 

In this paper we contribute to these efforts by studying NM in the emission process of a quantum emitter or ``atom" in front of {\it one} mirror, a model that can be solved exactly under very general approximations \cite{milonni,zoller,tufa}. One of the strengths of the considered system ---as explained in more detail below--- is that the crucial parameters ruling the occurrence of NM effects have a clear physical interpretation. In particular, within the limits of validity of the model, our study clearly illustrates how the non-Markovianity of the atomic emission is affected by imposing simple boundary conditions on the radiation field. In this spirit it is worth recalling that, even in the light of modern NM measures, spontaneous emission of a single atom (in vacuum) embodies the paradigmatic Markovian open dynamics: the emitted radiation simply travels away from the atom, so that the latter has no chance to retrieve information about its past dynamics from the electromagn!
 etic field (i.e. the environment). A typical way to establish information backflow in this scenario is to impose a geometrical confinement of the field, for example by means of a high-finesse cavity.  The NM of an atom in a cavity is often analyzed by assuming an {\it effective} Lorentzian-shaped  spectral density (SD) centered at a cavity protected frequency \cite{petruccione, lossy1}, and in the strong coupling regime can result even in vacuum Rabi oscillations \cite{jc}, an indisputably non-Markovian phenomenon.
A well-known implementation of a cavity is a Fabry-Perot resonator, which features {\it a pair of mirrors}. There is no fundamental reason, yet, that prevents NM from taking place even with {\it only one} mirror. Rather, in this context, introducing a single mirror in front of an atom appears to be the {\it minimum} geometrical confinement enabling the emergence of NM. Thus, from this viewpoint a simple atom-mirror setup -- otherwise termed an {\it atom in a half-cavity} -- can be regarded as a more fundamental system than a cavity to clarify how NM arises in the interaction of matter and geometrically confined light. Specifically, we will focus on a two-level atom where the emitted radiation is constrained to travel along a semi-infinite one-dimensional (1D) waveguide featuring a linear photonic dispersion relation. The finite end of the waveguide behaves as a perfect mirror, forcing part of the emitted light to return to the atom; one may expect such {\it feedback} mechanism to allow for information backflow, hence NM. Also, the finite time taken by a carrier photon to perform a round trip between atom and mirror ({\it time delay} $t_d$) should reasonably behave as an environmental memory time and hence as a key parameter to the occurrence of NM. The restriction to 1D geometry, while certainly a theoretical convenience, also ensures that a significant fraction of emitted light must return to the atom, which intuitively should enhance interference phenomena. These are ruled by the interplay between a {\it phase} parameter $\phi$, related to the atom-mirror optical path for a carrier photon, and the dimensionless parameter $\gamma t_d$, that is, the time delay rescaled by the spontaneous emission rate $\gamma$. We wonder if and how such interference affects NM. We find it remarkable that the occurrence of NM can be investigated as a function of quantities with a {clear physical interpretation}.

This paper is organized as follows. In Section \ref{open-dyn}, we briefly review the model under investigation, focussing the open system dynamics that the atom undergoes when it emits in vacuum. In Section \ref{NM-measure}, we tackle the problem of employing a reliable criterion to witness NM in the system under study and select a valid NM measure for a rigorous quantification. We explain our choice to use the measure in \rref\cite{tore}. The central findings of this work are then presented in Section \ref{central}, where we analize the occurrence of NM effects as a function of the two parameters $\gamma t_d$ and $\phi$ mentioned above. Special emphasis is given to the characterization of the threshold separating the Markovian and non-Markovian regions in the corresponding parameters space. We summarize our results and deliver some final comments in Section \ref{concl}. Further details on the treatment of the atom-mirror dynamics are given in Appendix A.

\section{Short review of the model} \label{open-dyn}
\begin{figure}
\begin{center}
\includegraphics[width=0.95\linewidth]{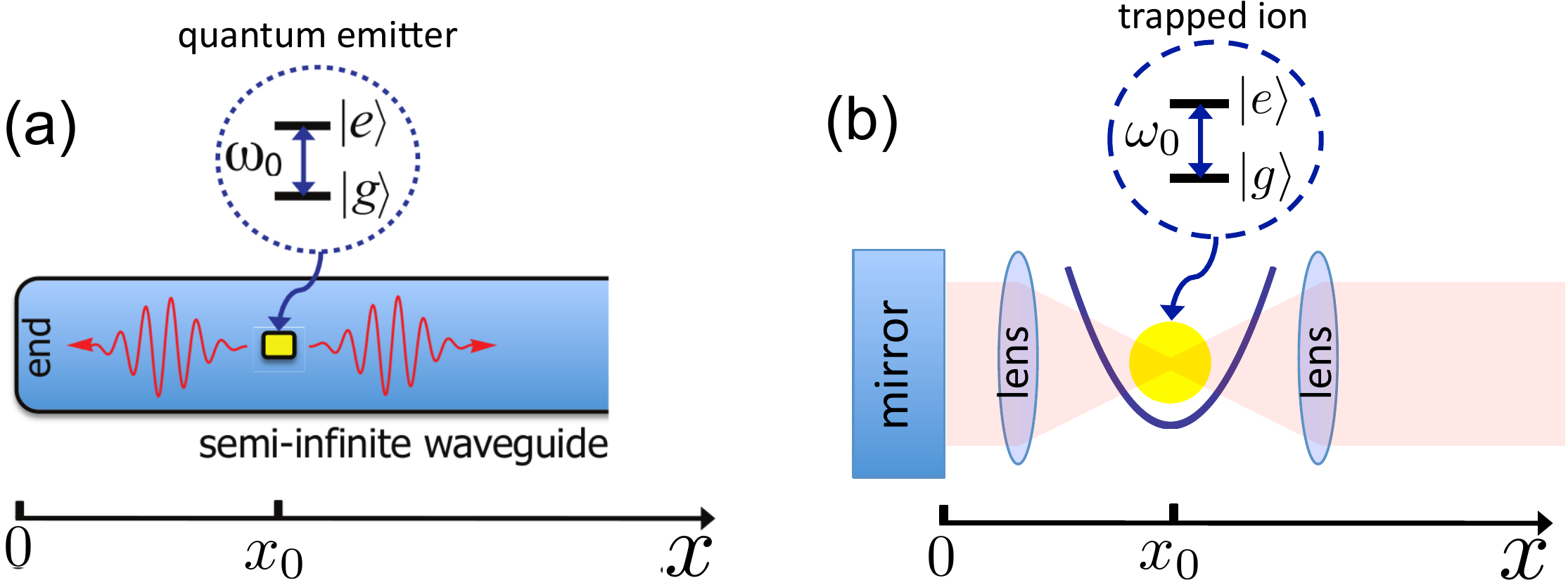}\end{center}\vspace{-.3cm}
\caption{(Color online) Two possible implementations of the model. (a) Semi-infinite waveguide, whose only end (behaving as a perfect mirror) lies at $x\ug0$, coupled to a two-level quantum emitter, such as a quantum dot, at $x\ug x_0$. (b) Free-space implementation featuring a trapped super-cool ion (quantum emitter), a real mirror and high-numerical-aperture lenses. \label{Fig1}}
\end{figure} 
The model we consider [see \fig1(a)] comprises a semi-infinite 1D photonic waveguide lying along the positive $x$-axis, containing a two-level quantum emitter (atom) placed at $x\ug x_0$. The waveguide termination at $x=0$ is assumed to behave as a perfect mirror, imposing a hard-wall boundary condition on the field. Several experimental implementations of the model are possible, involving a variety of quantum emitters embedded in several types of waveguides  (see {\it e.g.} Refs.~\cite{pc,fibers,wallraf,nws,gerard,delft,plasmons} and \rref \cite{sorensen} for a more comprehensive list). As shown in \fig1(b), a free-space implementation is viable as well along the lines of \rref\cite{free-space}. This makes use of a trapped ion, a standard mirror and a pair of high-numerical-aperture lenses. We remark that the 1D geometry is an idealization, and that the model could be refined by assuming the presence of external field modes into which the atom can decay \cite{zoller,tufa}.

The ground and excited states of the atom are denoted by $|g\rangle$ and $|e\rangle$ respectively, with energy separation $\omega_0$  ($\hbar\ug1$). The waveguide supports a continuum of electromagnetic modes, each with associated wave vector $k$ and frequency $\omega_k$. We assume that a linear dispersion relation of the form $\omega_k\simeq\omega_0+\upsilon(k-k_0)$, where $\upsilon$ is the photon group velocity and $k_0$ the carrier wave vector ($\omega_{k_0}\ug\omega_0$), is valid for a sufficiently broad band of modes around the atomic frequency $\omega_0$. 

The atom's emission process was first studied in the 80s \cite{milonni} and, more recently, revisited and extended in \rrefs \cite{zoller,tufa}. For the purposes of this work, it will be sufficient to recall the essential results allowing us to describe the reduced dynamics of the atom (the field being initially in the vacuum state). For more details we refer the reader to Appendix A and Refs.~\cite{milonni,zoller,tufa}. If the reduced state of the atom in the basis $\{\ket e,\ket g\}$ at time $t\!=\!0$ is 
\begin{equation}
\rho_0=\left(\begin{array}{cc}
\rho_{gg}&\rho_{ge}\\
\rho_{eg}&  \rho_{ee}
\end{array}\right)\label{rho0},
\end{equation}
with $\rho_{gg}\piu \rho_{ee}\ug1$ and $\rho_{eg}\ug\rho_{ge}^*$, it can be shown that at a later time $t$ its state reads
\begin{equation}
\rho(t)=\left(\begin{array}{cc}
\rho_{gg}+(1-|\varepsilon(t)|^2)\rho_{ee}&\varepsilon^*(t)\rho_{ge}\\
\varepsilon(t)\rho_{eg}& |\varepsilon(t)|^2 \rho_{ee}
\end{array}\right).\label{map}
\end{equation}
Here, $\varepsilon(t)$ is the probability amplitude to find the atom in state $|e\rangle$ at time $t$ when $\rho_0\ug|e\rangle\langle e |$. In a frame rotating at the atomic frequency $\omega_0$, the amplitude $\varepsilon(t)$ obeys the {\it delay differential equation}
\begin{align}
	\dot\varepsilon(t)\ug -\tfrac{\gamma}{2}\,\varepsilon(t)+\tfrac{\gamma}{2}\, {\rm e}^{i\phi}\varepsilon(t-t_d)\theta(t-t_d)\,\,,\label{DDE}
\end{align}
Where $t_d\ug 2x_0/\upsilon$ is the time delay [time taken by a photon to travel from the atom to the waveguide end and back, see \fig1(a)] , $\theta(t)$ is the Heaviside step function, and the phase $\phi\ug 2k_0x_0$ is the optical length for a carrier photon, corresponding to twice the atom-mirror path [in our convention, the $\pi$ phase shift due to reflection is taken into account by the different signs of the two terms on the righthand side of Eq.~\eqref{DDE}].
The crucial assumption in deriving Eq.~\eqref{DDE} is that the linearization of the waveguide dispersion relation has to be valid in a band of frequencies broader than the atomic width $\gamma$ and the inverse of the delay time $t_d^{-1}$ (see Appendix A). Note that this may still allow to have delay times much shorter than the spontaneous emission lifetime, i.e., $\gamma t_d\ll1$. 

Eqs.~\eqref{map} and (\ref{DDE}) fully determine the open dynamics of the atom. Note that the first term on the righthand side is associated with standard spontaneous emission. Instead, the feedback introduced by the presence of the mirror is represented by the second term. This is proportional to $\theta(t-t_d)$ meaning that, as expected, the atom undergoes standard spontaneous emission up to time $t\ug t_d$. After this, light emitted in the past can interfere with radiation emitted in the present as well as interact with the atomic dipole moment (i.e., excitation amplitude). Such interference process is witnessed by the phase factor $e^{i \phi}$ and, in general, can dramatically affect the dynamics. In particular, it can inhibit the full de-excitation of the atom for $\phi\ug2n\pi$ ($n$ integer), and in the regime $\gamma t_d\ll1$ it essentially prevents spontaneous emission altogether \cite{tufa}. 

Finally, we recall that the solution of \eq(\ref{DDE}) can be written as \cite{zoller,tufa} 
\begin{equation}\label{eps}
\varepsilon(t)={\rm e}^{\meno\frac{\gamma}{2}t}\sum_{n}\frac{1}{n!}\left(\tfrac{\gamma}{2}{\rm e}^{i\phi+\frac{\gamma}{2}t_d}\right)^n(t-nt_d)^n\theta(t-nt_d)\,.
\end{equation}

\section{Measuring quantum non-Markovianity}  \label{NM-measure}

As discussed in the Introduction, a number of NM measures have been proposed. A known issue is that, in general, such indicators are not equivalent and cases can be found where one of them vanishes while another one does not \cite{breuer}. A further problem is that their calculation is typically quite involved and may require optimization procedures. Such hurdles, yet, are mostly avoided in our case. The dynamical map in the form described by \eq(\ref{map}) can indeed be recognized as an {\it amplitude damping channel}. This type of channel for the atomic dynamics also arises in the case of the Jaynes-Cummings model and for an atom coupled to a lossy cavity with a Lorentzian spectral density \cite{petruccione}. In all these cases, a reliable criterion to test occurrence of NM can be expressed as \cite{breuer,lossy1}
\begin{equation}\label{criterion}
\frac{d|\varepsilon(t)|}{dt}<0\,\,\,\,\forall t>0\,\,\,\,\Leftrightarrow\,\,\,\,{\rm the\,\,system\,\,is\,\,Markovian}\,.
\end{equation}
In equivalent words, if $|\varepsilon(t)|$ (in fact, the atomic average energy) grows at some stage of time evolution (even though it may eventually fade away) then the dynamics is non-Markovian (and vice versa). This criterion relies on the demonstrable property \cite{breuer} that any open dynamics of the form (\ref {map}) is divisible if and only if $d |\varepsilon(t)|/dt\!\le\!0$ at any time, where indivisibility is recognized as a major trait of NM \cite{breuer}. Moreover, for this type of dynamics, relevant and in general non-equivalent measures of NM -- such as those in \rrefs \cite{rivas,BLP,tore} -- vanish iff condition (\ref{criterion}) holds. In our specific case, using \eq(\ref{DDE}) and the fact that the derivatives of $|\varepsilon(t)|$ and $|\varepsilon(t)|^2$ have the same sign,
criterion (\ref{criterion}) is equivalent to the condition \cite{nota}
\begin{equation}
\frac{d}{dt}|\varepsilon(t)|^2=-\gamma|\varepsilon(t)|^2\piu\gamma\,{\rm Re}\!\left[e^{i \phi}\varepsilon(t- t_d)\varepsilon^*(t)\right]\le0\,\,\,\,\,\,\,\,\forall\,t\geq t_d\,.\,\label{criterion2}
\end{equation}

While the study of conditions (\ref{criterion}) and \eqref{criterion2} is sufficient to distinguish between Markovian and non-Markovian regimes, \eg for assessing the existence of a threshold in parameter space separating the two, a specific measure has to be chosen in order to quantify NM. In this work, we adopt the recently introduced geometric measure of NM \cite{tore}. In general, this is defined as
\begin{equation}\label{meas}
\mathcal N\ug\frac{1}{V(0)}\int_{\frac{dV(t)}{dt}>0}\! \frac{dV(t)}{dt}\,,
\end{equation}
where $V(t)$ is the volume of accessible states of the open system $S$ under study at time $t$. The underlying idea is to view the dynamical map (associated with the considered open dynamics) as an affine transformation on the state space of $S$, which in the case of a qubit is the unit ball (the Bloch sphere). In this framework, a system is defined Markovian iff such volume can only shrink with time. This happens, in particular, with Lindbladian dynamical maps. For our dynamical map [Eq.~(\ref{map})], it is easily shown that the measure reads \cite{tore}
\begin{equation}
\mathcal N\ug\int_{\frac{d|\varepsilon(t)|^4}{dt}>0}\! \frac{d|\varepsilon(t)|^4}{dt}\,.\label{measure}
\end{equation}
Since $|\varepsilon(t)|^4$ is a monotonic function of $|\varepsilon(t)|$, \eq(\ref{measure}) enjoys a particularly straightforward connection with criterion (\ref{criterion}), making it a natural choice for our purposes. We stress, however, that this is an arbitrary choice since, as anticipated, the qualitative predictions on NM are mostly measure-independent for the present dynamics.

\section{Occurrence of non-Markovianity} \label{central}

Before explicitly computing $\mathcal N$, some general expectations on the emergence of NM can be formulated based on \eqs(\ref{DDE}), (\ref{criterion}) and (\ref{criterion2}). For the sake of clearness, we split the content of this section in three subsections corresponding to the regimes of negligible, very large and intermediate values of the rescaled delay time $\gamma t_d$, respectively. The last subsection in fact deals with the general case, hence reducing to the other two regimes in the limits $\gamma t_d\!\simeq\!0$ and $\gamma t_d\!\gg\!1$, respectively.
\begin{figure}
\begin{center}
\includegraphics[width=1\linewidth]{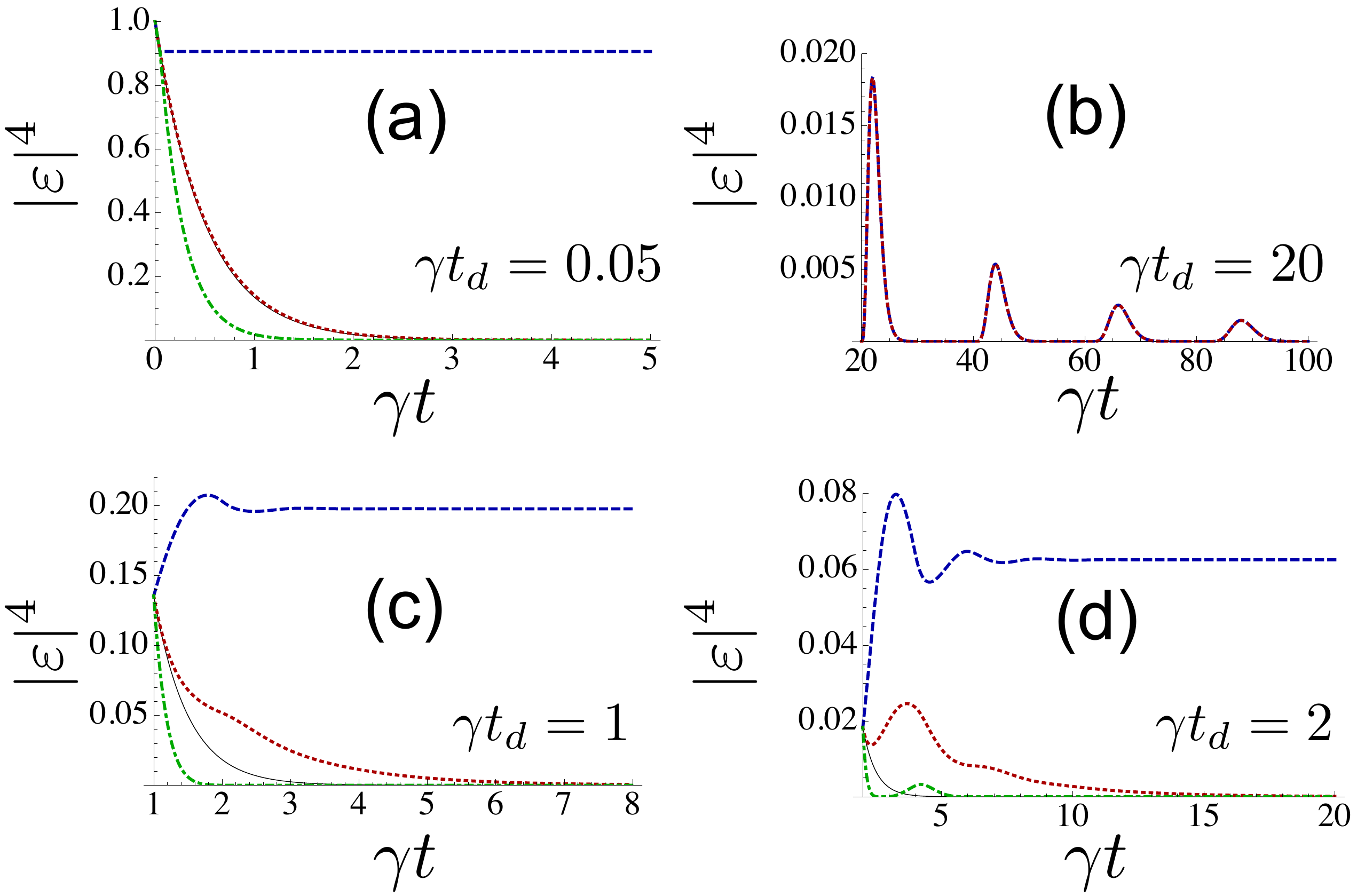}\end{center}\vspace{-.3cm}
\caption{(Color online) $|\varepsilon|^4$ against time (in units of $\gamma^{-1}$) for $\gamma t_d\ug0.05$ (a), $\gamma t_d\ug20$ (b), $\gamma t_d\ug1$ (c) and $\gamma t_d\ug2$ (d) and for $\phi\ug 0$ (blue dashed line), $\phi\ug \pi/2$ (red dotted) and $\phi\ug \pi$ (green dot-dashed). The black solid line corresponds to the case that the mirror is absent (standard spontaneous emission). At times $t\!\le\!t_d$,  it is always $\varepsilon(t)\ug e^{-\gamma/2t}$: in \figs(b), (c) and (d) we therefore plot only the behavior for $t\!>\!t_d$. \label{Fig2}}
\end{figure} 

\subsection{$\gamma t_d$ negligible: Lindbladian regime}\label{gammatausmall}
{\noindent}In the regime where $\gamma t_d$ is negligible \eq(\ref{DDE}) can be approximated as
\begin{align}
	\dot\varepsilon(t)\simeq \tfrac{\gamma}{2}\, ({\rm e}^{i\phi}-1)\varepsilon(t)\,\,,\label{DDEsmall}
\end{align}
and thus becomes local in time with time-independent coefficients. The corresponding behavior of $|\varepsilon(t)|$ and any power of it is clearly an exponential decay, as shown in \fig2(a) for $|\varepsilon(t)|^4$ with $\gamma t_d\ug0.05$. The dynamics therefore reduces to a Lindbladian one, hence {\it Markovian} \cite{zoller,tufa}. In such limit, the mirror feedback does not induce any NM, although -- depending on $\phi$ -- it can strongly affect the effective spontaneous emission rate, which can even be arbitrarily small for $\phi$ approaching $2n\pi$, in line with our previous discussion, or double for $\phi\ug n\pi$. Importantly, one has to single out the special case $\phi\ug2n\pi$, where a bound atom-photon state is formed regardless of the value of $\gamma t_d$ \cite{tufa}. As we show in Section \ref{intermediate} below, the dynamics for $\phi\ug 2n\pi$ is NM regardless of $\gamma t_d$.

\subsection{$\gamma t_d\gg1$: interference-free non-Markovian regime}\label{gammataularge}
The opposite regime takes place for $\gamma t_d\!\gg\!1$, which means that the time delay is far longer than the characteristic spontaneous emission time in absence of the mirror. The fraction of light emitted towards the mirror and then reflected back returns to the atom when this has already decayed to the ground state (and the light emitted in the opposite direction has fully departed). Such reflected light is reabsorbed by the atom and then emitted again in either direction and so on. As a consequence, in the regime $\gamma t_d\!\gg\!1$, $|\varepsilon(t)|^4$ exhibits successive spikes of decreasing height as shown in \fig2(b). Such behavior occurs independently of $\phi$ since, owing to the long retardation time, back-reflected light cannot recombine with light emitted towards the free end of the waveguide and no interference takes place.
Criterion (\ref{criterion}) thus entails that in this regime the dynamics is certainly {\it non-Markovian}. To compute the corresponding $\mathcal N$ [\cf\eq(\ref{measure})] we note that, as discussed in \rref\cite{zoller}, in the limit $\gamma t_d\!\gg\!1$  $\varepsilon(t)$ reduces in each interval to the last non-zero term of sum (\ref{eps}). Therefore, in each time interval $[m t_d,(m\!+\!1)t_d]$ ($m$ is a positive integer)
\begin{equation}\label{eps4}
|\varepsilon(t)|^4\simeq\left[\frac{\left(\tfrac{\gamma}{2}{\rm e}^{\frac{\gamma}{2}t_d}\right)^{m}}{m!}\right]^4\!{\rm e}^{\meno2\gamma t}(t-mt_d)^{4m}\,,
\end{equation}
which is explicitly independent of $\phi$. It is immediate to prove that the time derivative of this function is positive within the subinterval $[mt_d,mt_d\piu 2 m/\gamma]$, which is in agreement with the behavior in \fig2(b). Applying now \eq(\ref{measure}), we find
\begin{eqnarray}\label{N1}
\mathcal{ N}_{\gamma t_d\gg1}&\ug&\sum_{m=1}^\infty \left(\frac{\left(\tfrac{\gamma}{2}{\rm e}^{\frac{\gamma}{2}t_d}\right)^{m}}{m!}\right)^4 \left.\,|\varepsilon(t)|^4\phantom{\frac12}\!\!\!\right|_{m t_d}^{m t_d+2m/\gamma}\nonumber\\
&\ug&\sum_{m=1}^\infty\frac{ m^{4m}{\rm e}^{-4m}}{(m!)^4}\,,
\end{eqnarray}
where the convergence of the series is ensured by Stirling's approximation formula $n!\!\simeq\! n^n{\rm e}^{-n}\sqrt{2\pi n}$. Indeed, the summand in Eq.~\eqref{N1} asymptotically approaches $(2\pi m)^{-2}$. A numerical evaluation of the series provides $\mathcal{N}_{\gamma t_d\gg1}\!\simeq\!0.033$.

\subsection{General case: intermediate values of $\gamma t_d$}\label{intermediate}
Given the behavior in the limiting cases illustrated above, it is now interesting to investigate whether as $\gamma t_d$ grows from zero the system suddenly enters the non-Markovian regime or, instead, there is a finite threshold to trespass. If so, how does this threshold depend on $\phi$? Moreover, we wonder if the degree of NM as given by \eq(\ref{N1}) is the maximum possible or, instead, $\mathcal N$ can be higher at lower values of $\gamma t_d$ (due to interference effects, we may expect that the answer to this question depends on the phase $\phi$). The regime of intermediate values of $\gamma t_d$ features quite a rich physics with a variety of possible behaviors, as can be seen from \figs2(c) and (d) for two different values of $\gamma t_d$.

Although exact, the solution (\ref{eps}) of \eq(\ref{DDE}) is unfortunately complicated enough to prevent either $\mathcal N$ or even the mere NM condition (\ref{criterion}) from being worked out in a compact analytical form. We have therefore carried out a numerical computation of $\mathcal N$ by tabulating $|\varepsilon(t)|^4$ at the nodes of a time-axis mesh. Next, it was checked that the outcomes were stable with respect to the number of mesh points and the length of the overall simulated interval. 
\begin{figure}
\begin{center}
\includegraphics[width=0.85\linewidth]{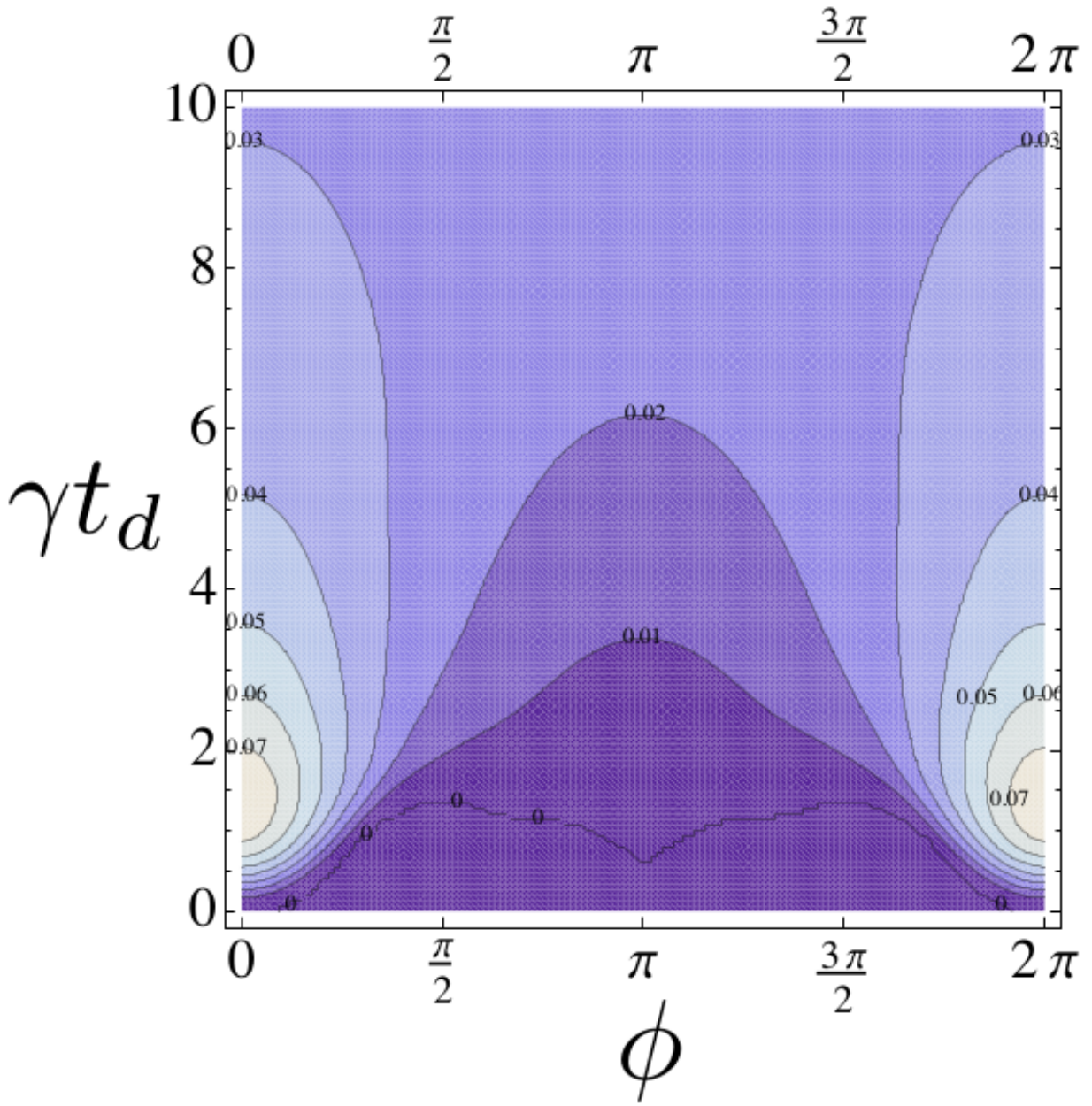}
\end{center}\vspace{-.3cm}
\caption{(Color online) Contour plot of $\mathcal N$ \vs $\phi$ and $\gamma t_d$. $\mathcal N$ is periodic with respect to $\phi$ with period $2\pi$. \label{Fig3}}
\end{figure} 
\fig3 shows a contour plot of $\mathcal N$ as a function of $\phi$ and $\gamma t_d$. The considered range of the phase $\phi$ is $[0,2\pi]$ due to the periodicity of the exponential. To begin our analysis of \fig3, we first observe that, as expected, $\mathcal N\!=\!0$ if $\gamma t_d\!\ll\!1$ \cite{nota-exception} (regime of negligible $\gamma t_d$, see Subsection \ref{gammatausmall}). On the other hand, as $\gamma t_d$ grows (regime of very large $\gamma t_d$) $\mathcal N$ converges to $\mathcal {N}_{\gamma t_d\gg1}\!\simeq\!0.03$ regardless of $\phi$, in line with the discussion related to \eq(\ref{N1}). As predicted, such asymptotic value is independent of $\phi$, which is witnessed by the fact that as $\gamma t_d$ grows the profile of $\mathcal N$ becomes more and more flat with respect to $\phi$. For a set value of $\gamma t_d$, the maximum of $\mathcal N$ is numerically found at $\phi\ug2n\pi$ and its minimum at $\phi\ug(2n\piu1)\pi$, where $n\!\ge\!0$ is an integer nu!
 mber. Such values of the phase shift correspond to the atom sitting at a {\it node} and {\it antinode}, respectively, of the field mode of wave vector $k_0$, that is, the mode resonant with the atomic transition [the sine function in \eq(\ref{H}), for this particular mode, can be recast as $\sin(\phi/2)$]. This might appear as counter-intuitive since NM is usually expected to increase with the effective atom-field coupling, which in turn grows with the field amplitude at the atom's location. However, it must be kept in mind that considerations about NM are typically model-dependent. For our system, a reasonable interpretation of these results can be given as follows. We note that the parameter $\phi$ encodes crucial information about the interference properties of the carrier wave-vector $k_0$, around which we expect to find most of the emitted radiation. More specifically, a carrier photon acquires a phase $\phi\piu\pi$ in a round trip between atom and mirror (the term $\pi$ due to mirror reflection). Thus, when $\phi\ug2n\pi$, the reflected field will return to the atom with an overall phase $\pi$ relative to the radiation that has been emitted towards the free end of the waveguide, resulting in destructive interference between the two. This effectively slows down the emission process, so that part of the emitted light can be expected to remain in the atom-mirror interspace for a significant time, which favours the occurrence of multiple re-absorptions [these bring about NM in virtue of \eq(\ref{criterion})]. Setting instead $\phi\ug n\pi$ (antinode), the interference between the reflected field and the freshly emitted one becomes constructive, thus enhancing the emission of radiation in the direction opposite the mirror. Obviously the latter is unable to re-excite the atom, which results in a reduced NM compared to the former situation. The difference between the two regimes, hence the gap between the maximum and minimum of $\mathcal N$ (see \fig3), becomes negligible as $\gamma t_d$ becomes very lar!
 ge. This can be understood by noting that, in such regime, the photon coming back from the mirror does not encounter any field with which it can interfere (as the atom will have decayed to the ground state well before a round-trip time). Equivalently, one might explain this by interpreting $\gamma t_d\gg1$ as the regime in which the ``bandwidth" $\gamma$ is large, compared to the characteristic frequency $1/t_d$: as $\gamma$ is increased the fraction of light at the carrier wave vector $k_0$ thus becomes less dominant in determining the behaviour of the emitted light. 
\begin{figure}
\begin{center}
\includegraphics[width=0.85\linewidth]{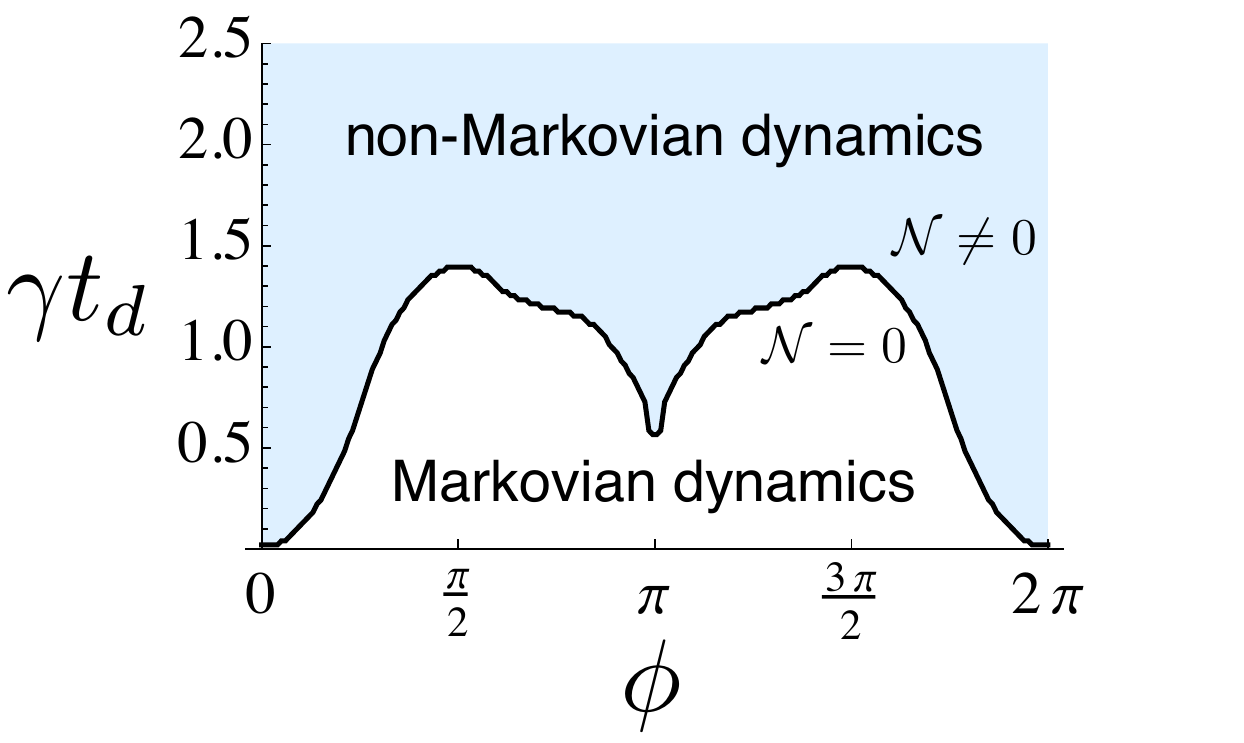}
\end{center}\vspace{-.3cm}
\caption{(Color online) Markovian and non-Markovian regions on the $\phi$-$\gamma t_d$ plane ($\mathcal N\!\neq\!0$ on the shaded area). The solid line represents the separation threshold.   \label{Fig4}}
\end{figure} 

A close inspection of \fig3 reveals the existence of a finite region on the $\phi$-$\gamma t_d$ plane within which the system exhibits a {\it Markovian} behavior, i.e., vanishing $\mathcal N$. The shape of such Markovianity region can be better appreciated in \fig4. For fixed $\phi$, one can find a finite {\it threshold} with respect to $\gamma t_d$ separating the Markovian and non-Markovian regime. The height of such threshold ranges from $\gamma t_d\ug0$ (for $\phi\ug0)$ to over $\gamma t_d\!\simeq\!1.4$ (for $\phi\ug\pi/2$). This indicates that, aside from the special point $\phi\ug0$, for fixed $\gamma$, $k_0$ and $\upsilon$ the mirror needs to lie far enough from the atom in order for the system to exhibit NM. Hence, when $\gamma t_d$ grows from zero the system in general does not enter suddenly the NM region.  Loosely speaking, the system can behave in a memoryless fashion even well beyond the Linbladian regime occurring for $\gamma t_d\!\simeq\!0$. The appearance of N!
 M thresholds in parameter space has been demonstrated in a number of systems such as in \rref \cite{lossy1,lossy2,pinja}. Interestingly, \fig4 shows the occurrence of thresholds even with respect to the phase shift $\phi$ for a fixed value $\gamma t_d$ (provided that this is lower than the threshold maximum). Owing to the discussed periodicity in $\phi$, this means that a continuous increase of $\phi$ makes the system cross in succession interspersed regions of Markovian and non-Markovian behavior. Interestingly, this can be achieved in practice by continuously detuning the atom's frequency (which is routinely attained through local fields) since this is easily seen to be equivalent to a change in $\phi$ \cite{tufa} (provided that the group velocity does not vary significantly in the explored frequency range). 

{\noindent}Both in terms of maximum amount of NM and threshold height, our results show that the most non-Markovian effects are found for a phase $\phi\ug2n\pi$. As anticipated, it can be demonstrated \cite{tufa} that such value of $\phi$ enables the formation of an atom-photon {\it bound state} in the atom-mirror interspace. This is in line with recent works pointing out the connection between NM and bound system-bath states \cite{bound}. In our specific system, it can be shown \cite{tufa} that an atom-photon bound state of energy $\omega_0$ (hence it is a bound state {\it in the continuum} \cite{bic}) is formed for $\phi\ug2n\pi$ between the atom and mirror (i.e., the corresponding photon density is identically zero for $x\!>\!x_0$). As a consequence, when $\phi\ug2n\pi$ part of the atomic excitation remains trapped according to \cite{tufa}
\begin{eqnarray}
\varepsilon(t\!\to\!\infty)\ug \frac{1}{1+\tfrac{\gamma t_d}{2}}\,,\label{FVT}
\end{eqnarray} 
which corresponds to the overlap between the atom's excited state and such bound state. Note that the trapping is reduced for increasing $\gamma t_d$. On the other hand, from \eq(\ref{DDE}), $\varepsilon(t_d)\ug e^{-\frac{\gamma t_d}{2}}$ which is lower than $\varepsilon(t\!\!\to\!\!\infty)$ for any $\gamma t_d\!>\!0$. Hence, $|\varepsilon(t)|$ must necessarily increase at some time, which in the light of criterion (\ref{criterion}) proves that the system is always non-Markovian at this special value of the phase.

{\noindent}Adopting a standard viewpoint in the theory of open quantum systems, we further observe that the phase $\phi$ determines the position of the atomic frequency with respect to the spectral density (SD) of the ``photonic bath". For our model, the spectral density is simply proportional to the square of the atom-photon coupling \cite{petruccione}, and can be expressed as
\begin{equation}
J(\Delta)\ug\frac{\gamma}{\pi}\sin^2\!\left(\frac{t_d}{2} \Delta+\frac\phi2\right)\,,\label{Jw}
\end{equation}
where we have defined the atom-photon detuning $\Delta\!\equiv\!\omega\meno\omega_0$ and we used the identity $\upsilon(k\meno k_0)\ug\Delta$. This leads to interpreting $t_d$ as the parameter ruling the width of the SD: as $t_d$ grows, $J(\Delta)$ exhibits an increasingly oscillatory behaviour. In the limit $t_d\!\rightarrow\!0$ and for fixed $\gamma$, the SD becomes flat which results in a Lindbladian dynamics (see Subsection \ref{gammatausmall} and \rref\cite{gardiner}). At the same time, the behaviour of $J(\Delta)$ around resonance is decided by $\phi$. Interestingly, such discussion allows for a natural comparison between our atomic dynamics in a single-mirror setup and that occurring in a lossy cavity featuring a Lorentzian SD. In the latter case, the dynamics is characterized by two dimensionless parameters: these are $\gamma\lambda^{-1}$, where $\lambda$ measures the SD's width, $\gamma$ being again the spontaneous emission rate in the ``flat spectrum" limit, and  $!
 \delta/\gamma$, with $\delta$ the atomic detuning with respect to the cavity protected frequency (at which the maximum of the SD occurs). Significantly, despite major differences between the two systems, also the lossy cavity model exhibits NM thresholds with respect to both the width parameter and detuning \cite{tore,lossy1,lossy2}. While a lossy cavity has long been considered the paradigmatic system in which to investigate the emergence of NM effects, we find it interesting that a similarly rich structure can occur even with a single mirror.  We also observe that the strength of NM effects, as quantified by Eq.~(\ref{meas}), appears comparable in the two models, if the Lorentzian SD is taken in the ``bad cavity limit" $\gamma\lambda^{-1}\!\lesssim\!5$. This can be appreciated by comparing \fig3 and the results in \rref\cite{tore}. A significant difference between the two models, however, is the fact that the single-mirror setup features an absolute maximum ${\cal N}_{\sf!
  max}\!\simeq\!0.07$ as a function of the model parameters, wh!
 ile for the lossy cavity ${\cal N}$ can be made arbitrarily large by increasing the cavity quality factor.

\section{Conclusions} \label{concl}
We have studied the occurrence of NM in the emission process of an atom coupled to a one-dimensional field, in the presence of a single mirror which imposes a hard-wall boundary condition on the latter. In general, the resulting open dynamics of the atom exhibits a non-exponential behaviour with a rich structure. Adopting the non-divisibility of time evolution as the chosen definition of NM, and a corresponding NM quantifier proposed in \rref\cite{tore}, we have studied the strength of NM effects in our system as a function of the two effective parameters characterizing the model: the rescaled round-trip time $\gamma t_d$ and the phase $\phi$. 
While analytical results have been provided in the limiting cases $\gamma t_d\ll1$ and $\gamma t_d\gg1$, a numerical approach has been adopted for the intermediate regime $\gamma t_d\sim{\cal O}(1)$. Remarkably, a finite region in parameter space can be identified where no NM occurs, its boundary defining a NM threshold. For any fixed value of $\gamma t_d$, the maximum NM is found at $\phi\ug2n\pi$, where a bound atom-photon state is formed. Interestingly, finite Markovian thresholds occur with respect to both the SD width parameter and atomic detuning, a structure which is also exhibited in the open dynamics of an atom in a lossy cavity with Lorentzian spectral density. A deeper and more rigorous insight into the relationship between the NM effects in such cavity model and those occurring in the half-cavity treated here can be gained by introducing a second imperfect mirror in the latter model. The lossy cavity dynamics is then obtained in the limit of negligible time delay!
 s \cite{milonni} (a cavity model featuring non-null time delays has been investigated in \rref\cite{tureci,ref3}). The analysis of NM for such a two-mirror model, which can be regarded as an {\it ab initio} -- instead of phenomenological -- description of a lossy cavity, is currently under investigation \cite{tufa2}.\\

\subsection*{Acknowledgements}
{TT and MSK acknowledge support from the NPRP
4-554-1-084 from Qatar National Research Fund. We are grateful to P.~Haikka,Y.~Pijlo, B.~Garraway, M.~Paternostro and S.~Lorenzo for useful discussions.}
\appendix
\section{Short review of the model} \label{model} 
An atom in a half-cavity is conveniently described through a Quantum Electrodynamics (QED)-like approach under the usual rotating-wave approximation (RWA), where the mirror enforces a boundary condition on the field. Here we review some essential features of this model \cite{milonni,zoller,tufa}. Denoting the annihilation (creation) bosonic operator of the waveguide field as  $\hat{a}_k$ ($\hat{a}^\dagger_k$), the Hamiltonian reads
\begin{equation}
	\hat{H}\ug\omega_0\kebra{e}{e}\piu\!\!\int_0^{k_c}\!\!\!{\rm d}k\;\omega_k\hat a^\dagger_k\hat a_k\piu  \!\sqrt{\frac{\gamma \upsilon}{\pi}}\!\!\int_0^{k_c}\!\!\!{\rm d}k\,\sin kx_0\!\left(\hat\sigma_+\hat a_k\piu\textrm{H.c.}\right),\label{H}
\end{equation}
where $\hat\sigma_+\ug\hat\sigma_-^\dag\ug\kebra{e}{g}$, $k_c$ stands for a cut-off wave vector and $\gamma$ is the atomic spontaneous emission rate (if the waveguide were infinite).  In \eq(\ref{H}), note that the coupling strength between the atom and the $k$th mode is $\propto\!\sin{kx_0}$, which stems from the constraint that the field vanishes at the mirror location $x\ug0$ (hard-wall boundary condition). As specified in the main text, we are concerned with the reduced dynamics of the atom when the field is initially in the vacuum state $|0\rangle$. The total number of excitations is conserved since $[\hat{H},\kebra{e}{e}\piu\int\!{\rm d}k\;\hat a_k^\dagger \hat a_k]\ug0$. Note that the state $|g\rangle|0\rangle$ does not evolve in time since it is an eigenstate of $\hat H$ (with zero eigenvalue). On the other hand, $|e\rangle|0\rangle$ evolves in a superposition of all possible single-excitation atom-field states. For such initial condition, the joint atom-field system!
  evolves to the (globally) pure state
\begin{equation}
\ket{\Psi(t)}\ug\varepsilon(t)\ket e\!\ket 0\piu\ket{g}\!\!\int\!\!{\rm d}k\; \varphi(k,t) \,a^\dagger_k\!\ket{0},\label{generic-state}
\end{equation}
where $\varepsilon(0)\ug1$ and $\varphi(k,t)$ is the field amplitude in $k$-space. From this, it is immediate to derive Eq.~\eqref{map} for the evolution of the atomic reduced state.
To work out $\varepsilon(t)$, one makes use of the the time-dependent Schr\"{o}dinger equation $i{|\dot\Psi}(t)\rangle\ug\hat H|\Psi(t)\rangle$, which results in a system of differential equations for $\varepsilon(t)$ and $\varphi(k,t)$ \cite{zoller,tufa}. Two approximations are then made. (i) It is assumed that the photon dispersion relation can be linearized as $\omega_k\!\simeq\! \omega_0\piu \upsilon (k\meno k_0)$, where $\upsilon\ug\left.{{\rm d}\omega}/{{\rm d}k}\right|_{k\ug k_0}$ is the photon group velocity  and $k_0$ is the wave vector corresponding to the atomic frequency, i.e., $\omega_{\kappa_0}\ug \omega_0$.  (ii) The integral bounds are approximated as $\int_0^{k_c} {\rm d}k\!\simeq\!\int_{-\infty}^\infty {\rm d}k$. These routine approximations \cite{gardiner}, together with the RWA mentioned earlier, rely on the fact that only a narrow range of wave vectors around $k\!\ug \! k_0$ is expected to give a significant contribution to the dynamics. Since the range !
 of frequencies involved in the the atomic dynamics is ruled by the two parameters $\gamma$ and $t_d$, we deduce that the linearization of the waveguide dispersion relation is a good approximation in a band of frequencies broader than the atomic width $\gamma$ as well as the inverse of the delay time $t_d^{-1}$. Clearly, a further requirement is that the time delay $t_d$ should be much larger than the optical period $\omega_0^{-1}$, in order to avoid the breakdown of the RWA. In specific implementations of the model, these assumptions have to be checked {\it a posteriory} for consistency. Once we set in a rotating reference frame such that $\varepsilon(t)\!\to\!\varepsilon(t) {\rm e}^{-i\omega_0t},\varphi(k,t)\!\to\!\varphi(k,t){\rm e}^{-i\omega_0t}$ and the field variables are expressed in terms of the atomic excitation amplitude\cite{zoller,tufa}, we end up with Eq.~\eqref{DDE} in the main text.

\end{document}